\documentclass[aps,prl,twocolumn,showpacs,preprintnumbers,amsmath,amssymb]{revtex4} 
 
\usepackage{graphicx}
\usepackage{dcolumn}
\usepackage{bm}
\def\lsim{\raise0.3ex\hbox{$<$\kern-0.75em\raise-1.1ex\hbox{$\sim$}}} 
\def\gsim{\raise0.3ex\hbox{$>$\kern-0.75em\raise-1.1ex\hbox{$\sim$}}}

\def\Teff{T_{\rm eff}} 
\def\psat{p_{\rm sat}} 
\def\avpt{\langle p_{\rm T}\rangle} 
 
\def\av#1{\langle#1\rangle}

\def\Teff{T_{\rm eff}} 
 
\def\Tdec{T_{\rm dec}} 
 
\newcommand{\beq}{\begin{equation}} 
\newcommand{\eeq}{\end{equation}} 
\newcommand{\bea}{\begin{eqnarray}} 
\newcommand{\eea}{\end{eqnarray}}

\begin{document} 
\preprint{HIP-2002-26/TH}

\title{Dependence of hadron spectra on \\ 
decoupling  temperature and resonance contributions} 
 
\author{ 
K.J. Eskola$^{\rm a,b}$, H. Niemi$^{\rm a}$, P.V. Ruuskanen$^{\rm a,b}$, 
S.S. R\"as\"anen$^{\rm a}$ } 
\email{kari.eskola, harri.niemi, vesa.ruuskanen, sami.rasanen@phys.jyu.fi} 
\affiliation{$^{\rm a}$Department of Physics, 
P.O.Box 35, FIN-40014 University of Jyv\"askyl\"a, Finland\\ 
$^{\rm b}$Helsinki Institute of Physics, 
P.O.Box 64, FIN-00014 University of Helsinki, Finland}

\date{24.06.2002, revised 13.12.2002}

\begin{abstract} 
 
Using equilibrium hydrodynamics with initial conditions for the energy and 
net baryon number densities from the perturbative QCD + saturation model, a 
good simultaneous description of the measured pion, kaon and (anti)proton 
spectra in central Au+Au collisions at $\sqrt s=130\,A$GeV is found with a 
single decoupling temperature  $\Tdec=150\dots160$ MeV. The interplay between 
the resonance content of the EoS and the development of the transverse flow 
leads to inverse slopes and $\langle p_T\rangle$ of hadrons which 
increase with decreasing $\Tdec$. The origin of this result is discussed. 
 
\end{abstract}

 
\maketitle

{\em Introduction.} The amount of initially produced matter in 
ultrarelativistic heavy ion collisions at collider energies has been 
suggested to be controlled by gluon saturation 
\cite{SATURATION,EKRT,KL}.  For heavy nuclei, $A\sim 200$, the 
saturation scale which determines the dominant transverse momentum 
scale is expected to be $1\dots 2$ GeV, which can lead to quite large 
values for the initial transverse energy \cite{EKRT,ERRT,RAJU}.  For 
instance, the initial $dE_T/dy$ at $y\sim 0$ obtained from the 
perturbative QCD (pQCD) + saturation model \cite{EKRT} exceeds the 
value observed in central Au+Au collisions at Brookhaven RHIC by a 
factor $2.5 \dots 3$ \cite{EKRT,ERRT}. 
 
A mechanism to transfer transverse energy into the 
longitudinal motion is provided by the asymmetry of the initial 
collective motion of produced matter.  The observed rapidity 
distributions of final particles suggest that the matter is produced 
in a state of rapid longitudinal expansion \cite{Huovinen:1998tq}.  In 
the collective expansion energy is transferred through the work by 
pressure, $pdV$. Hence the strong initial expansion in longitudinal 
direction leads to a large transfer of energy from the transverse 
into the longitudinal motion. The space--time evolution of a 
dense matter with collective effects and with a QCD phase transition 
is describable in terms of relativistic hydrodynamics 
\cite{Teaney:2001av,PVR,HEINZ,Huovinen:2002fp}. 
 
Hadron spectra are often argued to be insensitive to the early stages 
of the collision.  This is not quite true for events with large 
multiplicities since secondary collisions will lead to transverse 
collective motion and correlations in the properties of particle 
spectra which are difficult to understand otherwise. 
A clear correlation resulting from radial flow is the mass dependence of 
slopes of transverse spectra.  For non-central collisions hydrodynamics 
predicts the experimentally observed elliptic flow \cite{PASIH}. 

As energy is transferred into longitudinal direction during the 
expansion, it is usually expected that the transverse spectra of final 
pions become steeper at lower decoupling temperatures.  However, since 
the number of hadrons and resonance states increases rapidly with mass, 
a considerable fraction of the energy of thermal matter can be in the 
form of heavy resonances.  When the total transverse energy of all 
hadrons decreases with the decreasing decoupling temperature, the energy 
released from the reduction of the number of resonances and heavy 
particles can lead to an increase of the slope of the spectra of the 
remaining final stable hadrons.  The energy release from the latent heat 
in a phase transition has a similar but smaller effect. In this work we 
study the details of these effects and the interplay between them by 
comparing the expansion of a massive pion gas with that of a hadron 
resonance gas. 
 
Slopes of hadron spectra from heavy ion experiments at Brookhaven AGS and 
CERN SPS, can be reproduced with a kinetic freeze-out temperature 
$\sim 120$ MeV \cite{Sollfrank_prc}. To reproduce the strange 
particle abundances in thermal models, chemical freeze-out temperatures 
higher than 120 MeV are needed \cite{REDLICH}. Also at RHIC, a higher 
decoupling temperature is supported by the heavy particle yields \cite{HEINZ}. 
In this note we argue that in central Au+Au collisions at 
$\sqrt s=130$~$A$GeV also the kinetic freeze-out takes place effectively at 
higher temperatures: solving the transverse expansion within boost invariant 
hydrodynamics with initial conditions for the energy and net baryon number 
densities from the pQCD + saturation model \cite{EKRT,ERRT}, 
we describe simultaneously the measured $\pi^\pm,\ K^\pm$ and $p(\bar p)$ 
spectra with a single decoupling temperature $T_{\rm dec}=150\dots160$~MeV.  
Similar observation has been made in Ref.~\cite{Florkowski} using a 
fireball parametrization. 
 
We describe briefly our framework to calculate the hadron spectra and then 
discuss how the hadron and resonance content in the Equation of State (EoS), 
and the phase transition affect the dependence of the slope of the 
spectra on the decoupling temperature.  Finally, the pion, kaon and 
(anti)proton spectra are compared with the experimental results from 
PHENIX~\cite{PHENIX_data,PHENIX_QM02,PHENIX_lambda}.

{\em The theoretical framework.} 
The calculational frame which we use has been discussed in detail in 
\cite{ERRT}. The initial particle production is calculated from the 
pQCD + saturation model based on lowest order pQCD interactions and a 
cut-off scale $\psat$ determined from a saturation condition for the 
final state minijets \cite{EKRT}. The $K$-factor 2.3 has been fixed on 
the basis of the next-to-leading order calculation of minijet transverse 
energy~\cite{ET}. 
 
The transverse energy weighted minijet cross section in a central 
rapidity unit, $\sigma\av{E_T}(\psat,\sqrt s, A,|y|\le 0.5)$, is the 
key quantity for determining the initial energy density at the time of 
formation of the matter.  As a new ingredient, we include the 
corresponding quantity for the initial net baryon number, 
$\sigma\langle N_B\rangle 
=(\sigma\langle N_q \rangle - \sigma\langle N_{\bar q} \rangle)/3$, 
where the computation of $\sigma\langle N_{q(\bar q)} \rangle $ is 
based on the flavour decomposition of the minijet cross sections in 
\cite{EK}. 
 
Below, we shall consider Au+Au collisions at $\sqrt s=130$~$A$GeV with 
a 5~\% centrality cut. As explained in \cite{ERRT}, this corresponds to 
central collisions of effective nuclei with $A_{\rm eff}=181$. The 
number of participants which we get for such collisions is 346, 
consistent with the PHENIX result $348\pm10$ \cite{PHENIX_data}.  For the 
computation of the initial densities, we obtain $\sigma\av{E_T}=67.0$ 
mbGeV, $\sigma\langle N_B\rangle=0.64$ mb and $\psat=1.06$ GeV. 
Thus, the initial $E_T=1750$~GeV and $N_B=16.7$ in the interval 
$|y|\le0.5$ \cite{ERRT}. 
 
The matter is assumed to be (approximately) thermalized at the time 
$\tau_0=1/\psat=0.19$ fm/$c$.  For central collisions, this approach 
has predicted successfully \cite{EKRT,PVR} the particle multiplicities 
at RHIC \cite{RHIC_mult} and also, when amended with a hydrodynamical 
description of the transverse expansion of produced matter, the 
transverse energies of final particles, $dE_T/dy$, within 10\dots 20~\% 
\cite{ERRT}. 
 
The high temperature phase of our EoS is QGP with a bag constant $B$ and 
the low temperature phase a hadron resonance gas (HRG) with hadrons and 
hadron resonances up to $M=2$~GeV and a repulsive mean field 
among the hadrons \cite{Huovinen:1998tq}.  The values of the bag 
constant $B$ and the mean field strength $K$ are so chosen that the 
transition temperature is $T_c=165$ MeV. 
 
The thermal spectra of final stable hadrons are obtained by folding the 
thermal motion with the fluid motion using the Cooper and Frye prescription 
\cite{CooperandFry,PVR}.  The dependence of the spectra on the transverse 
flow velocity $v_r=\tanh y_r$ and on the decoupling temperature $\Tdec$ 
enters through the factors $(p_T/\Tdec)\cosh y_r$ and $(m_T/\Tdec)\sinh 
y_r$ as the arguments of modified Bessel functions \cite{PVR}. E.g., for 
pions with $p_T\gsim 1$ GeV, contributions from regions where $v_r\gsim0.5$ 
behave approximately as $\sim \exp(-p_T/e^{y_r}\Tdec)= \exp(-p_T/\Teff)$ 
with $\Teff=e^{y_r}\Tdec$. This shows explicitly how the inverse slope, 
$\Teff$, depends at large $p_T$ on the decoupling temperature and on the 
transverse flow rapidity $y_r$.  Due to the exponential dependence on 
$y_r$, a small increase in the flow can compensate the decrease in 
$\Tdec$. The net effect on $\Teff$, as the decoupling temperature 
and the transverse flow change, depends on the EoS and, in 
particular, on the assumed particle content of the hadron phase. 
 
To obtain the spectra of final stable hadrons, we first calculate the 
spectra of all hadrons and hadron resonances which are included in the 
EoS. Then the decay contributions from all resonance states are added 
to the spectra of stable hadrons. All strong and electromagnetic two 
and three body decay modes with known branching ratios are included 
\cite{PDG}.

%
%
\begin{table*}[htb] 
\hspace{0.0cm} 
\begin{tabular}{|c|c|c|c|c|c|c|c|c|c|c|c|c|c|} 
\hline 
\hline 
 & \multicolumn{2}{|c|}{\rule[-0.2cm]{0cm}{0.6cm}PG (a)} 
 & \multicolumn{2}{|c|}{HRG (b)} 
 & \multicolumn{2}{|c|}{PG+QGP (c)} 
 & \multicolumn{2}{|c|}{HRG+QGP (d)} 
 & \multicolumn{5}{|c|}{QGP+HRG+RD (e)} 
\\ 
\hline 
\hline 
 $\Tdec$& 
{\Large$\frac{dN^{\pi^+}}{dy}$\rule[-0.4cm]{0cm}{1.2cm}} & 
$\langle p_T^{\pi^+}\rangle$\rule[-0.4cm]{0cm}{1.2cm}& 
{\Large$\frac{dN^{\pi^+}}{dy}$\rule[-0.4cm]{0cm}{1.2cm}} & 
$\langle p_T^{\pi^+}\rangle$\rule[-0.4cm]{0cm}{1.2cm}& 
{\Large$\frac{dN^{\pi^+}}{dy}$\rule[-0.4cm]{0cm}{1.2cm}} & 
$\langle p_T^{\pi^+}\rangle$\rule[-0.4cm]{0cm}{1.2cm}& 
{\Large$\frac{dN^{\pi^+}}{dy}$\rule[-0.4cm]{0cm}{1.2cm}} & 
$\langle p_T^{\pi^+}\rangle$\rule[-0.4cm]{0cm}{1.2cm}& 
{\Large$\frac{dN^{\pi^+}}{dy}$\rule[-0.4cm]{0cm}{1.2cm}} & 
$\langle p_T^{\pi^+}\rangle$\rule[-0.4cm]{0cm}{1.2cm}& 
{\Large$\frac{dN}{dy}$\rule[-0.4cm]{0cm}{1.2cm}} & 
{\Large$\frac{dE_T}{dy}$\rule[-0.4cm]{0cm}{1.2cm}} & 
{\Large$\frac{dE_T}{d\eta}$\rule[-0.4cm]{0cm}{1.2cm}} 
\\ 
\hline 
 160 &  211 & 0.93 & 
        111 & 0.54 & 
        392 & 0.55 & 
         88 & 0.51 & 
        258 & 0.44 & 
        1122 & 701 & 570\\ 
\hline 
 130 &  204 & 0.96 & 
        203 & 0.57 & 
        394 & 0.56 & 
        159 & 0.54 & 
        262 & 0.51 & 
        1041 & 668 & 588\\ 
\hline 
 100 &  192 & 1.02 & 
        319 & 0.61 & 
        374 & 0.61 & 
        238 & 0.57 & 
        272 & 0.56 & 
        959 & 630  & 584\\ 
\hline 
\end{tabular} 
\caption{ Multiplicity $dN^{\pi^+}/dy$ and average transverse momentum 
$\langle p_T^{\pi^+}\rangle/$GeV of positive pions for two 
different decoupling temperatures $\Teff/$MeV and the same four 
EoS as in Fig.~1.  Resonance decays (RD) have been added in the case 
(e). The  multiplicity $dN/dy$ and the transverse energies $(dE_T/dy)/$GeV 
and $(dE_T/d\eta)/$GeV for all particles are shown for the case (e). 
} 
\label{table1} 
\end{table*}

\begin{figure}[bht] 
\vspace{-0.3cm} 
\hspace{-0.65cm} 
\includegraphics[height=7.3cm]{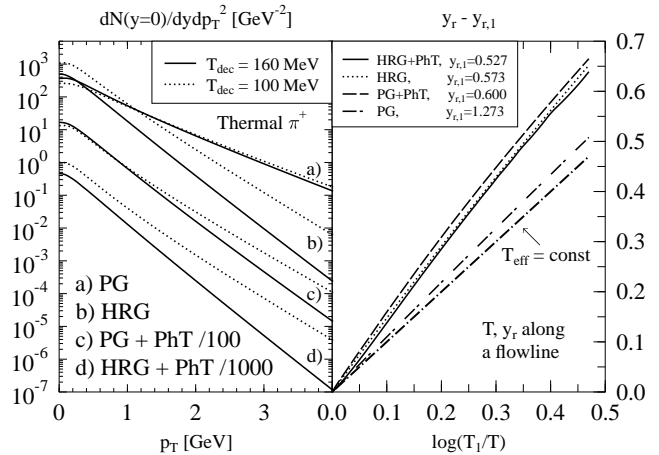} 
\label{inicond} 
\vspace{-0.9cm} 
\caption{\protect\small {\bf Left:} Transverse spectra of 
thermal $\pi^+$ for two decoupling temperatures $\Tdec$ in the 
four cases studied; decay pions are not included for (b) and (d). 
The spectra for (c) and (d) have been scaled. 
{\bf Right:} The change of the transverse flow rapidity $y_r$ as a 
function of temperature along a flow line 
starting at $r=5~$ fm at $\tau_0$. The flow rapidities $y_{r,1}$ at 
$T_1=160$~MeV are indicated. On the dashed--dotted line the effective 
temperature $\Teff=e^{y_r}\Tdec$ would remain constant. 
} 
\end{figure} 

{\em The effect of resonances and phase transition on $\Tdec$ dependence 
of spectra.} Before comparing our results with data, we try to clarify 
the $\Tdec$ dependence of final hadron spectra by comparing the spectra 
of thermal pions at $\Tdec=160$ and 100 MeV in four different cases: 
(a) massive pion gas (PG) without phase transition (PhT), 
(b) hadron resonance gas (HRG) without PhT and 
(c) PG with PhT to QGP, 
(d) HRG with PhT to QGP. 
To make the differences in the effects on flow more transparent, we 
first consider only thermal pions also in cases (b) and (d).  The 
spectra of positive pions are shown in the left panel of Fig.~1 for each 
case.  Except in the case (a) the slopes in the interval $1.5\lsim 
p_T\lsim 4$ GeV change considerably.  Pions in this part of the spectrum 
come mainly from matter with large transverse velocity where $T_{\rm 
eff}=e^{y_r}\Tdec$ holds. 

In the expansion the temperature of the matter drops and the transverse 
velocity increases. These changes have opposite effects on the slope. 
They nearly cancel if the growth of the flow in terms of an increase in 
$y_r$ satisfies $\Delta y_r=y_r-y_{r,1}=\log(T_1/T)$ when the temperature 
drops from $T_1$ to $T$. In the right panel of Fig.~1  we plot $\Delta y_r$ 
as a function of $\log(T_1/T)$ between $T_1=160$~MeV and $T=100$~MeV for a 
fluid element moving along a flow line which initially starts at $r=5$~fm. 
This fluid element belongs to the region which dominates the tail of the 
spectra.  We see that for PG without PhT, the case (a) with the small change 
of slope (cf. Fig. 1, left), the curve is close to the dashed-dotted line 
for $e^{y_r}\Tdec=const$. For the other cases the increase in flow with the 
decrease in temperature is stronger and the changes in $\Delta y$ are 
comparable. This indicates a similar relative increase in $\Teff$, the 
inverse slope of the spectra, in the cases (b)-(d). 
 
The development of transverse flow is fastest for PG with large pressure 
gradients and slower for the other cases with softer EoS.  In the cases 
(b) and (d) with HRG, a considerable amount of energy is stored in the 
heavy hadron and resonance states.  As the matter expands, the release 
of energy from these states slows down the decrease in temperature 
leading to a stronger growth of the flow, as a function of decreasing 
temperature, than in the PG of the case (a).  In the case (c), the 
latent heat released at the strong first order PhT leads to a similar 
behavior of the $y_r(T)$ as in the case (b) and (d).  The reason why the 
change of slopes between (b) and (d) is smaller than between (a) and (c) 
is the weaker phase transition in the case (d).  The strength of the PhT 
is controlled by $s_{Q}/s_{H}$, the ratio of entropy densities of the 
QGP and hadron phase at $T_c$ which is $\simeq 15$ for the case (c) and 
$\simeq 3$ for (d).  If the ratio in the case (c) is artificially forced 
to that in (d) by decreasing the number of degrees of freedom in the 
QGP, the changes in the pion spectra from (a) to (c) and (b) to (d) 
become approximately the same.  We conclude, that for the realistic 
EoS with HRG and QGP, the change in the slope of thermal pions, is 
caused mainly by heavy hadrons and resonances in the EoS and to much 
lesser degree by the phase transition. 
 
The values of the average transverse momentum $\langle p_T\rangle$ and 
multiplicities of positive thermal pions are collected to the Table 
\ref{table1}.  In addition, the column (e) shows the same quantities for 
HRG+QGP when also the pions from the resonance decays (RD) are included. 
The results show that $\langle p_T\rangle$'s of thermal pions increase 
$\sim10$ \% when $T_{\rm dec}$ drops from 160 to 100 MeV in all the 
cases (a)-(d). However, the change is $\simeq 21$ \% in (e), where 66~\% 
of the pions come from the decays at $T_{\rm dec}=160$~MeV (cf. 
Table~\ref{table2}). In this case the low-$p_T$ part of the pion spectrum, 
that gives the dominant contribution to multiplicity and $\langle 
p_T\rangle$, is filled by decay pions from slowly moving resonances.  As 
$\Tdec$ decreases the fraction of resonances drops increasing the growth of 
the average transverse momentum. The same behaviour is seen in the case (b), 
if decay pions are included. The last three columns of Table~\ref{table1} 
show the total multiplicity $dN/dy$ of all particles and their total 
transverse energies $dE_T/dy$ and $dE_T/d\eta$, where $E_T=E\sin\theta$ 
\cite{ERRT}.

\begin{figure}[h!bt] 
\label{paneli} 
\vspace{-0.cm}\hspace{-0.3cm} 
   \includegraphics[height=5.6cm]{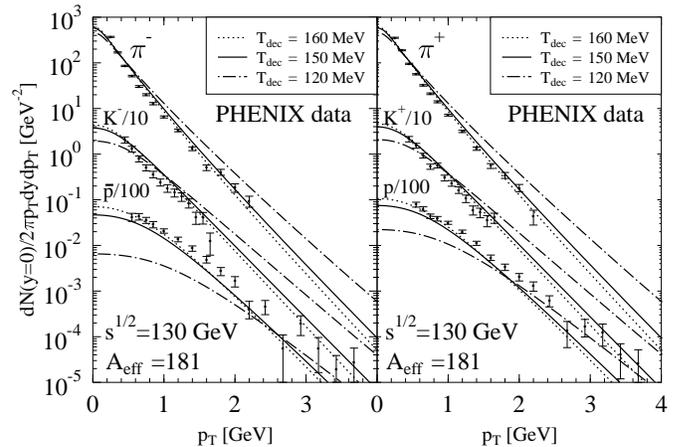} 
\vspace{-.3cm} 
\caption{\protect\small Transverse spectra of charged hadrons in Au+Au 
collisions at $\sqrt s=130$~$A$GeV with a 5 \% centrality cut.  The data 
is from PHENIX \cite{PHENIX_data} and the curves are our results for 
$\Tdec=$~120, 150 and 160 MeV.} 
\vspace{-0.cm} 
\end{figure} 

{\em Comparison with data.} 
Next we demonstrate that (i) using the initial conditions from the 
pQCD + saturation model and (ii) the EoS HRG+QGP with large number of hadron 
states in the low temperature phase, it is possible to find a {\em single} 
decoupling temperature which reproduces both the measured abundances of 
pions, kaons and (anti)protons and the shapes of their spectra. 
 
%
\begin{table}[htb] 
\hspace{0.1cm} 
\begin{tabular}{|c|c|c|c|c|c|c|c|c|c||c|} 
\hline 
\hline 
 &\multicolumn{3}{|c|}{positive pions } 
 &\multicolumn{3}{|c|}{positive kaons } 
 &\multicolumn{4}{|c|}{protons }  \\ 
\hline 
\hline 
 {\Large$\frac{T_{\rm dec}}{\rm MeV}$} & 
{\Large$\frac{dN}{dy}$\rule[-0.4cm]{0cm}{1.2cm}} & 
 $F$ & 
 $\avpt$ & 
{\Large$\frac{dN}{dy}$\rule[-0.4cm]{0cm}{1.2cm}} & 
 $F$ & 
 $\avpt$ & 
{\Large$\frac{dN}{dy}$\rule[-0.4cm]{0cm}{1.2cm}} & 
 $F$ & 
 $\avpt$ & 
 $\bar p/ p$ \\ 
\hline 
 160 & 258 & 0.34 & 0.44 & 52.1 & 0.50 & 0.61 & 24.5 & 0.31 & 0.80 & 0.69 \\ 
\hline 
 150 & 259 & 0.42 & 0.46 & 51.2 & 0.56 & 0.65 & 20.2 & 0.36 & 0.86 & 0.65 \\ 
\hline 
 120 & 265 & 0.71 & 0.53 & 40.7 & 0.75 & 0.78 & 9.79 & 0.55 & 1.08 & 0.34 \\ 
\hline 
\end{tabular} 
\caption{ The multiplicity $dN/dy$, the fraction $F$ of thermal 
particles in the multiplicity, and the average transverse momentum 
$\avpt$ in GeV for positive pions, kaons and protons calculated at 
three decoupling temperatures. The last column is the antiproton-to-proton 
ratio $\bar p/p\equiv (dN^{\bar p}/dy)/(dN^p/dy)$. 
} 
\label{table2} 
\end{table}

Transverse spectra of pions, kaons and (anti)protons, calculated at 
$\Tdec=120$, 150 and 160 MeV, are shown in Fig.~2 with the data 
from PHENIX \cite{PHENIX_data}.  Spectra of negative (positive) 
particles are in the left (right) panel.  We find the overall 
agreement between the data and the calculations at $\Tdec= 150\dots160$~MeV 
very satisfactory, considering that we have not adjusted parameters other 
than the decoupling temperature. A $\sim 30$ \% feed-down from hyperons is 
included in the $p(\bar p)$ data \cite{PHENIX_data} but not in the 
calculation. Especially, we would like to draw attention to the $\Tdec$ 
dependence of the slopes: at $\Tdec\sim 150\dots160$~MeV the slopes agree 
with the slopes of the data but at 120 MeV all spectra are too shallow. 
 
In Table~\ref{table2} we list the multiplicities and the thermal 
fractions $F$ for positive pions, kaons and protons, also at $\Tdec=160$, 
150 and 120 MeV. Our results on the multiplicities agree within the 
experimental errors with the measured values \cite{PHENIX_data}. For 
pions the thermal fraction $F$ increases strongly because the number of 
heavier resonances drops fast. This stronger suppression of heavier 
states at lower temperature is the sole source for the relative change 
between the multiplicities of pions and heavier particles; it does not 
depend on the flow at all. 
 
The widening of spectra with decreasing $\Tdec$ is also seen in
Table~\ref{table2} as an increase in the average transverse momenta.
For pions and kaons our results at $\Tdec=150\dots160$~MeV are $\sim
10$~\% larger than the measured values and slightly outside the
experimental error bars \cite{PHENIX_data}. The $dE_T/d\eta$ which we
obtain with $\Tdec=150\dots160$~MeV agrees within 8 \% with the latest
PHENIX result \cite{PHENIX_QM02}, when the difference in the
definitions of $E_T$ is taken into account.

PHENIX has also reported $p$ $(\bar p)$ yields corrected for hyperon 
feed-down, $dN/dy=19.3\pm0.6$ $(13.7\pm0.7)$ \cite{PHENIX_lambda}. Our 
results in Table~\ref{table2} for $\Tdec=150$ MeV agree with the 
measurement.  With $\Tdec=150$ (160) MeV we also obtain 
$\bar\Lambda/\Lambda=0.70$ (0.74), again consistent with the PHENIX 
result $0.75\pm0.09$ \cite{PHENIX_lambda}. Note that the deviation of 
$\bar p/p$ and $\bar\Lambda/\Lambda$ from unity is due to the 
net baryon number content of the initial matter at saturation. 
 
We have not performed a $\chi^2$ fit for finding the best value for 
$\Tdec$ since our aim is to see how well we can describe the main 
features of the data while keeping our framework as simple as possible: 
pQCD + saturation to calculate the initial state, EoS to govern the 
expansion, and $\Tdec$ to specify the freeze-out.  Also all the input 
parameters with the exception of $\Tdec$ are based on our previous studies 
\cite{ERRT,Huovinen:1998tq,EK} and we have not tried to tune them here.

{\em Summary}. Using a hydrodynamic framework, we have studied the 
effect of the EoS and the decoupling temperature on the particle spectra 
and multiplicities. We have demonstrated an interesting and significant 
interdependence between the hadron content of the EoS, the decoupling 
temperature and the strength of the flow. Taking the initial conditions 
from the pQCD + saturation model \cite{EKRT,ERRT}, we have calculated the 
transverse spectra and the multiplicities of pions, kaons and (anti)protons 
at the RHIC energy $\sqrt{s}=130$~$A$GeV with a 5~\% centrality cut in 
gold--gold collisions. The agreement between our results with 
$\Tdec=150\dots160$ MeV and the data measured by the PHENIX Collaboration 
\cite{PHENIX_data,PHENIX_lambda} is very good both for the shape and the 
normalization of the spectra.

\begin{acknowledgments} 
We thank P. Huovinen and C. Salgado for discussions. 
Financial support from the 
Academy of Finland, project no. 50338 is gratefully acknowledged. 
\end{acknowledgments}

\end{document}